\documentclass[aps,prltightenlines,superscriptaddress,nofootinbib,amsmath,amssymb,reprint]{revtex4-2}
 
\usepackage{lineno} 
\usepackage{graphicx}
\usepackage{dcolumn}
\usepackage{bm}
 \usepackage{verbatim}
 \usepackage{xcolor}

\RequirePackage{color}

\begin{document}

\preprint{AIP/123-QED}

\title[Sample title]{Measurement of coherent elastic neutrino nucleus scattering on germanium by COHERENT}

\newcommand{\USDdesc}{\affiliation{Department of Physics, University of South Dakota, Vermillion, SD, 57069, USA}}
\newcommand{\SNUdesc}{\affiliation{Department of Physics and Astronomy, Seoul National University, Seoul, 08826, Korea}}
\newcommand{\FSUdesc}{\affiliation{Department of Physics, Florida State University, Tallahassee, FL, 32306, USA}}
\newcommand{\Dukedesc}{\affiliation{Department of Physics, Duke University, Durham, NC, 27708, USA}}
\newcommand{\TUNLdesc}{\affiliation{Triangle Universities Nuclear Laboratory, Durham, NC, 27708, USA}}
\newcommand{\Mephidesc}{\affiliation{National Research Nuclear University MEPhI (Moscow Engineering Physics Institute), Moscow, 115409, Russian Federation}}
\newcommand{\ITEPnewadesc}{\affiliation{National Research Center  ``Kurchatov Institute'' , Moscow, 123182, Russian Federation }}
\newcommand{\UTKdesc}{\affiliation{Department of Physics and Astronomy, University of Tennessee, Knoxville, TN, 37996, USA}}
\newcommand{\NCSUdesc}{\affiliation{Department of Physics, North Carolina State University, Raleigh, NC, 27695, USA}}
\newcommand{\Sandiadesc}{\affiliation{Sandia National Laboratories, Livermore, CA, 94550, USA}}
\newcommand{\UCASdesc}{\affiliation{University of Chinese Academy of Sciences, Beijing, 100049, China}}
\newcommand{\Tuftsdesc}{\affiliation{Department of Physics and Astronomy, Tufts University, Medford, MA, 02155, USA}}
\newcommand{\ORNLdesc}{\affiliation{Oak Ridge National Laboratory, Oak Ridge, TN, 37831, USA}}
\newcommand{\LANLdesc}{\affiliation{Los Alamos National Laboratory, Los Alamos, NM, 87545, USA}}
\newcommand{\CNLdesc}{\affiliation{Canadian Nuclear Laboratories Ltd, Chalk River, Ontario, K0J 1J0, Canada}}
\newcommand{\ICRRdesc}{\affiliation{University of Tokyo, Institute for Cosmic Ray Research, Kamioka, Gifu, 506-1205, Japan}}
\newcommand{\Okayamadesc}{\affiliation{Department of Physics, Okayama University, Okayama, Okayama, 700-8530, Japan}}
\newcommand{\CMUdesc}{\affiliation{Department of Physics, Carnegie Mellon University, Pittsburgh, PA, 15213, USA}}
\newcommand{\IUdesc}{\affiliation{Department of Physics, Indiana University, Bloomington, IN, 47405, USA}}
\newcommand{\VTdesc}{\affiliation{Center for Neutrino Physics, Virginia Tech, Blacksburg, VA, 24061, USA}}
\newcommand{\Hawaiidesc}{\affiliation{Department of Physics and Astronomy, University of Hawaii, Honolulu, HI, 96822, USA}}
\newcommand{\NCCUdesc}{\affiliation{Department of Mathematics and Physics, North Carolina Central University, Durham, NC, 27707, USA}}
\newcommand{\NCSUnucengdesc}{\affiliation{Department of Nuclear Engineering, North Carolina State University, Raleigh, NC, 27695, USA}}
\newcommand{\WJCdesc}{\affiliation{Washington \& Jefferson College, Washington, PA, 15301, USA}}
\newcommand{\Tokyodesc}{\affiliation{Department of Physics, University of Tokyo, Tokyo, 113-0033, Japan}}
\newcommand{\Kyotodesc}{\affiliation{Department of Physics, Kyoto University, Kyoto, Kyoto, 606-8502, Japan}}
\newcommand{\UFdesc}{\affiliation{Department of Physics, University of Florida, Gainesville, FL, 32611, USA}}
\newcommand{\Concorddesc}{\affiliation{Department of Physical and Environmental Sciences, Concord University, Athens, WV, 24712, USA}}
\newcommand{\SLACdesc}{\affiliation{SLAC National Accelerator Laboratory, Menlo Park, CA, 94025, USA}}
\newcommand{\Laurentiandesc}{\affiliation{Department of Physics, Laurentian University, Sudbury, Ontario, P3E 2C6, Canada}}
\author{M.~Adhikari}\USDdesc
\author{M.~Ahn}\SNUdesc
\author{D.~Amaya Matamoros}\FSUdesc
\author{P.S.~Barbeau}\Dukedesc\TUNLdesc
\author{V.~Belov}\Mephidesc\ITEPnewadesc
\author{I.~Bernardi}\UTKdesc
\author{C.~Bock}\USDdesc
\author{A.~Bolozdynya}\Mephidesc
\author{R.~Bouabid}\email{ryan.bouabid@duke.edu}\altaffiliation{Now at: Los Alamos National Laboratory, Los Alamos, NM, 87545, USA}\Dukedesc\TUNLdesc
\author{A.~Bracho}\Dukedesc\TUNLdesc
% ...

\date{\today}

\begin{abstract}
The COHERENT collaboration reports the most precise measurement of the coherent elastic neutrino-nucleus scattering cross section to date. This measurement was performed with COHERENT’s germanium detector array, Ge-Mini, at the Spallation Neutron Source at Oak Ridge National Laboratory. A cumulative exposure of \(4.68\times10^{22}\) protons on target yielded a total number of observed counts of \(124^{+14}_{-12}\) and a flux-averaged cross section of \(1.00 \pm  0.10\,\text{(statistical)} \pm 0.10\,\text{(systematic)}\) relative to the standard-model expectation of \(5.9\times10^{-39}\,\mathrm{cm}^2\). The well-understood energy and timing distributions of the neutrino source allow for independent measurements of muon- and electron-neutrino scattering rates. This information is used to improve constraints on non-standard neutrino interactions mediated by heavy particles.
\end{abstract}

\maketitle
\paragraph*{Introduction.---}
Coherent elastic neutrino nucleus scattering (CEvNS) is a low-energy, weak neutral-current interaction where a neutrino scatters off of a nucleus coherently. 
This coherence condition amplifies the differential scattering rate, scaling with the square of the nuclear weak charge.
The only detectable signal from this interaction is the nuclear recoil, at energies of order keV and below. 
CEvNS measurements require both low energy thresholds and low backgrounds to enable observations of the small neutrino-induced recoils. 
The interaction is precisely predicted by the standard model, making it an excellent probe of neutrino properties, nuclear structure, and new physics.

CEvNS was first proposed in 1974~\cite{Freedman:1973yd,Kopeliovich:1974mv} and received its first confirmation in 2017 by the COHERENT Collaboration~\cite{COHERENT:2017ipa}.
Since then, the study of CEvNS has grown into a broad experimental program boasting several observations from both terrestrial and astrophysical sources~\cite{COHERENT:CsIPRL,COHERENT:2020ArPRL,COHERENT:GePRL,CONUS:2025obx,XENON:2024ijk, PandaX:2024muv, LZ:2025igz}. 
Combined analyses of these measurements have demonstrated their complementary phenomenological impact~\cite{AtzoriCorona:2025xgj, Giunti:2024gec}. 
Of particular importance to future long-baseline neutrino experiments is the possibility of non-standard neutrino interactions (NSI) modifying the interpretation of upcoming oscillation data~\cite{Coloma:2015kiu}. 
While such interactions would appear as deviations from the clean standard-model predictions of CEvNS, existing CEvNS measurements are statistically limited in their ability to constrain these scenarios~\cite{Giunti:2019xpr}.

In pursuit of precision CEvNS measurements, the COHERENT Collaboration has deployed a suite of detectors to the Spallation Neutron Source (SNS) at Oak Ridge National Laboratory (ORNL)~\cite{COHERENT:2026ewu}.
The SNS is the world’s most intense stopped-pion neutrino source, producing a large, pulsed neutrino flux with well determined spectral characteristics in energy and time~\cite{COHERENT:flux}. 
The COHERENT detectors are located in “neutrino alley” (NA), a corridor in the basement of the target facility that provides excellent shielding from beam-produced neutrons and reduces cosmic ray backgrounds.

The COHERENT Ge-Mini detector is a germanium based CEvNS experiment that reported the first observation of CEvNS on germanium using data collected in 2023~\cite{COHERENT:GePRL}. 
Since that measurement, roughly three times the neutrino exposure has been collected.
In addition, refined analysis techniques have enabled a lower analysis threshold, thereby increasing the CEvNS signal acceptance, while the application of pulse shape discrimination has significantly reduced backgrounds. 
This Letter reports the results of the new data-taking campaign, yielding the most precise measurement of CEvNS to date and the collaboration’s first measurement limited by the dominant systematic uncertainty from the neutrino flux normalization~\cite{COHERENT:flux}. 
Using the unique timing structure of the SNS, this dataset is further used to set leading constraints on NSI induced by heavy mediators.

\paragraph*{Experiment.---}
The Ge-Mini detector consists of an array of p-type inverted coaxial point contact (ICPC) germanium detectors. 
The detectors exhibit low electronic noise, with measured pulser resolution of 150 eV\(_{\text{ee}}\) (electron-recoil equivalent energy) full-width half-maximum or better, enabling an analysis threshold of 500 eV\(_{\text{ee}}\). 
Regions of reduced electric field near the detector surfaces (on the order of mm) lead to modified charge-drift dynamics and characteristic pulse-shape differences, enabling discrimination between surface backgrounds and bulk interactions. 
The detector array is surrounded by graded passive shielding against environmental and beam-related radiation and an active muon veto to suppress cosmogenic backgrounds~\cite{COHERENT:GePRL}.

Ge-Mini is located \(19.2 \pm 0.1\)\,m from the SNS target, in the former location of the COHERENT CsI detector, where beam-related neutron backgrounds have been extensively characterized~\cite{COHERENT:CsIPRL}. 
This measurement used four ICPC detectors with a total bulk, active mass of \(8.53 \pm 0.08\)\,kg, as determined by the manufacturer Mirion Technologies. The active mass corresponds to the bulk volume of the diode, where energy depositions are not subjected to surface effects.
The uncertainty on the active mass is subdominant to the neutrino flux uncertainty.

During nominal operations, the SNS accelerates protons into a mercury target at a repetition rate of 60\,Hz. 
Throughout the data-taking campaign in 2025, the proton energy was 1.3 GeV, with approximately equal running periods of beam powers at 1.7\,MW and 1.8\,MW. 
The total number of beam spills during data taking was recorded and converted to an equivalent number of protons on target and corresponding neutrino exposure using Monte Carlo estimates of pion production~\cite{COHERENT:flux} giving an expected 0.37\(\pm\)0.04 neutrinos per proton.

\paragraph*{Data taking.---} 
The data-taking campaign reported here began on February 15\(^{\text{th}}\), 2025 and ended on May 27\(^{\text{th}}\), 2025. 
Detector energy signals were digitized at a sampling rate of 125 MHz for a total recorded waveform length of 176 \(\mu\)s.
During the collection of physics data, the digitization of the germanium waveforms was triggered by external timing signals, thereby enabling the collection of output signals that is independent of pulse amplitude.  
This experiment used two such external timing signals: one synchronized with protons on target (on-beam), and one delayed by 3.03 ms (off-beam) to sample steady-state backgrounds under identical detector conditions. 
In addition to the germanium waveforms, logic signals indicating preamplifier transistor resets were recorded continuously during data-taking. 

During scheduled weekly beam down-time, germanium detector outputs were digitized in a “self-triggered” mode to collect steady-state background data with approximately an order of magnitude larger statistics than externally triggered datasets.
This dataset determined the energy scale for the analysis, provided inputs for steady-state background modeling, and was used to validate the pulse shape discrimination.
All data reported here were collected after servicing and recommissioning following the first data-taking campaign, which led to improved stability and operating conditions.

\paragraph*{Data analysis.---}

\begin{figure}
    \centering
    \includegraphics[width=1.0\linewidth]{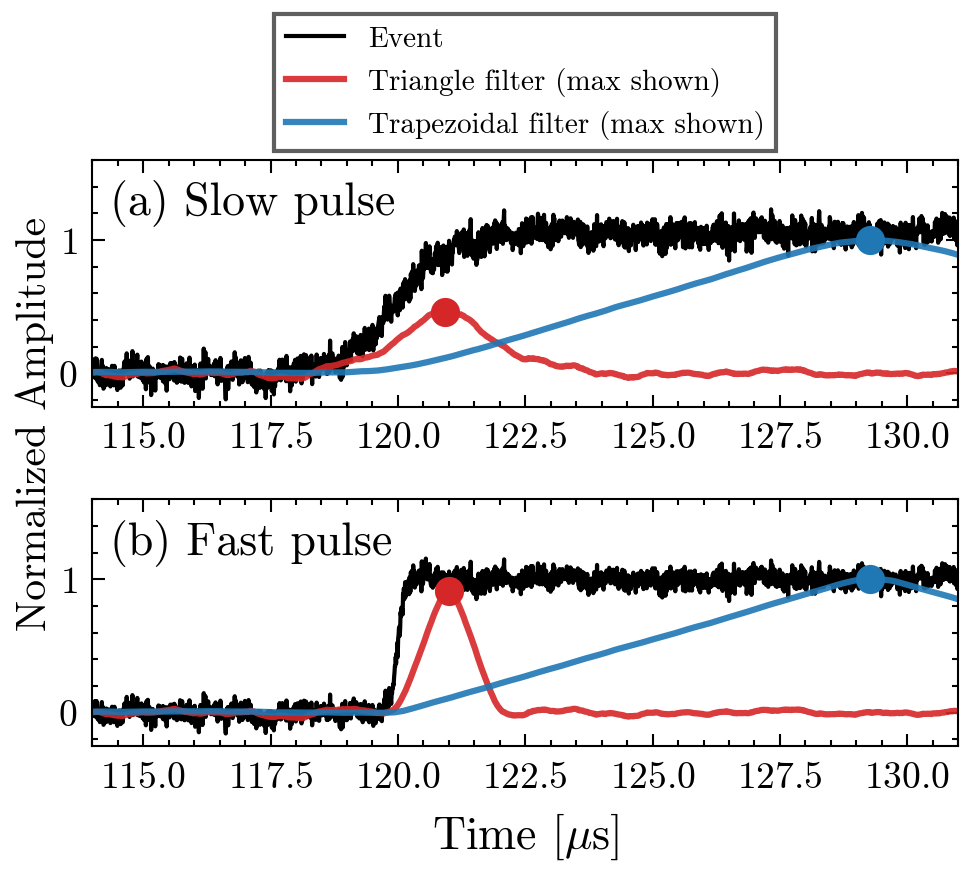}
    \caption{Example waveforms from the Ge-Mini detectors: (a) a surface-like event and (b) a signal-like event. The raw pulse (black) is overlaid with the output of the matched triangle filter (red) used to determine the pulse onset and the output of the trapezoidal filter (blue) used to reconstruct the pulse energy; circles indicate the maxima of each filter. The pulse shape parameter, defined as the ratio of the triangle-filter maximum to the trapezoidal-filter maximum, is close to unity for signal-like events and smaller for slow-rise surface-background events.}
    \label{fig:waveforms}
\end{figure}

The waveform reconstruction and cuts were developed using the background datasets during a blinded analysis. 
For each waveform, three features were extracted through offline digital signal processing: pulse amplitude, pulse onset, and a pulse shape parameter. 
The energy of an event, proportional to the pulse amplitude, is reconstructed through a trapezoidal filter~\cite{Radeka:1988kpb}, optimized for each detector based on the electronic noise (with rise times between 8 and 12\,\(\mu\)s and \(0.67\,\mu\mathrm{s}\) flat time). 
This method was demonstrated to reconstruct pulse amplitudes in an unbiased manner down to 0.2 keV\(_{\text{ee}}\). 
Pulse onset identification was one of the major limiting factors in determining the analysis energy threshold of 1.5 keV\(_{\text{ee}}\) in Ref.~\cite{COHERENT:GePRL}. 
In this analysis, the pulse onset is reconstructed using a matched triangle filter with 1~\(\mu\)s rise time for every detector. 
This method was demonstrated to have a pulse onset finding resolution better than 150 ns down to 0.5 keV\(_{\text{ee}}\). 
Finally, a pulse shape parameter was extracted from the waveforms to reject events with slow rise times, characteristic of surface backgrounds. 
In this analysis, the ratio of the maximum of the matched triangle filter and the maximum of the energy-optimized trapezoidal filter was used as the pulse shape parameter. 
An example of a signal-candidate event and a likely background surface event are shown in Fig.~\ref{fig:waveforms} with trapezoidal and triangle filters applied.

The response of the pulse shape parameter as a function of energy was characterized using simulated waveforms overlaid with measured noise.
The pulse-shape cut was chosen conservatively to ensure \(>\)99\% signal acceptance across the analysis energy range and was fixed prior to unblinding.
Reconstructed waveforms passing the pulse shape cut are considered if they are in the energy and onset region of interest (ROI): [0.5, 20]\,keV\(_{\text{ee}}\) and aligned in time between [-4, 40]\,\(\mu\)s with respect to the measured protons on target (and subsequent neutrino arrival) in the on-beam dataset, and the equivalent portion of the recorded waveforms in the off-beam dataset. 

Non-physical events, including microphonic excitations and digitizer artifacts, were rejected through a combination of waveform quality selections and time period removals described below. The overall waveform quality selection strategy and validation were similar to that in Ref.~\cite{COHERENT:GePRL}, though pathological waveforms specific to this data-taking period and the lower energy threshold required their own treatment.

The waveform-based cuts removed events from the analysis that were not consistent with the characteristics of energy depositions in the crystal. Spurious negative going pulses were identified and cut if the trapezoidal filter minimum was more than twice the amplitude of the maximum. 
Events that saturated the digitization dynamic range were also removed. 
Waveforms were rejected if the fractional difference between the baseline slope estimates computed over the first 40 \(\mu\)s and first 112 \(\mu\)s of the waveform exceeded 25\%. 
The waveform-based cuts described here were defined conservatively during the blinded analysis and validated by direct visual inspection of all candidate waveforms flagged by these selections, with no signal-like events rejected.

In this analysis, with the help of machine-learning clustering techniques, preamplifier resets in one detector were identified as the source of cross-talk in digitizer channels for other detectors.
This crosstalk was the origin of the events previously removed in Ref.~\cite{COHERENT:GePRL} through cuts on the trapezoidal filter minimum. 
Preamplifier resets and induced cross-talk events were effectively cut by excluding events from all detectors within a fixed 200~\(\mu\)s window around any reset, resulting in an approximately \(1\%\) dead-time. 
Because preamplifier resets are asynchronous with respect to beam timing, no signal bias or energy-dependent efficiency is introduced from this event selection.
Data collected during daily liquid nitrogen Dewar fills are excluded from analysis. Periods of increased noise due to microphonic activity were also removed. Events were excluded from the analysis if a coincident muon veto channel was triggered between [-16, 40]\,\(\mu\)s relative to the analysis window with reconstructed energy in the muon veto above an energy threshold optimized to reduce the dead-time induced by spurious coincidences. 

With the analysis techniques described above, a common low energy threshold of 0.5\,keV\(_{\text{ee}}\) was adopted across all detectors to ensure uniform treatment and robust background rejection.
The combination of conservative waveform-quality selections and the exclusion of known periods of high noise resulted in an overall dead-time that is independent of signal energy. After all data selection, the total dead-time was exactly measured for each detector and ranged between 11 and 13\%, similar to the performance reported in Ref.~\cite{COHERENT:GePRL}.

\paragraph*{Backgrounds.---}

\begin{figure*}[t]
  \centering
  \includegraphics[width=\textwidth]{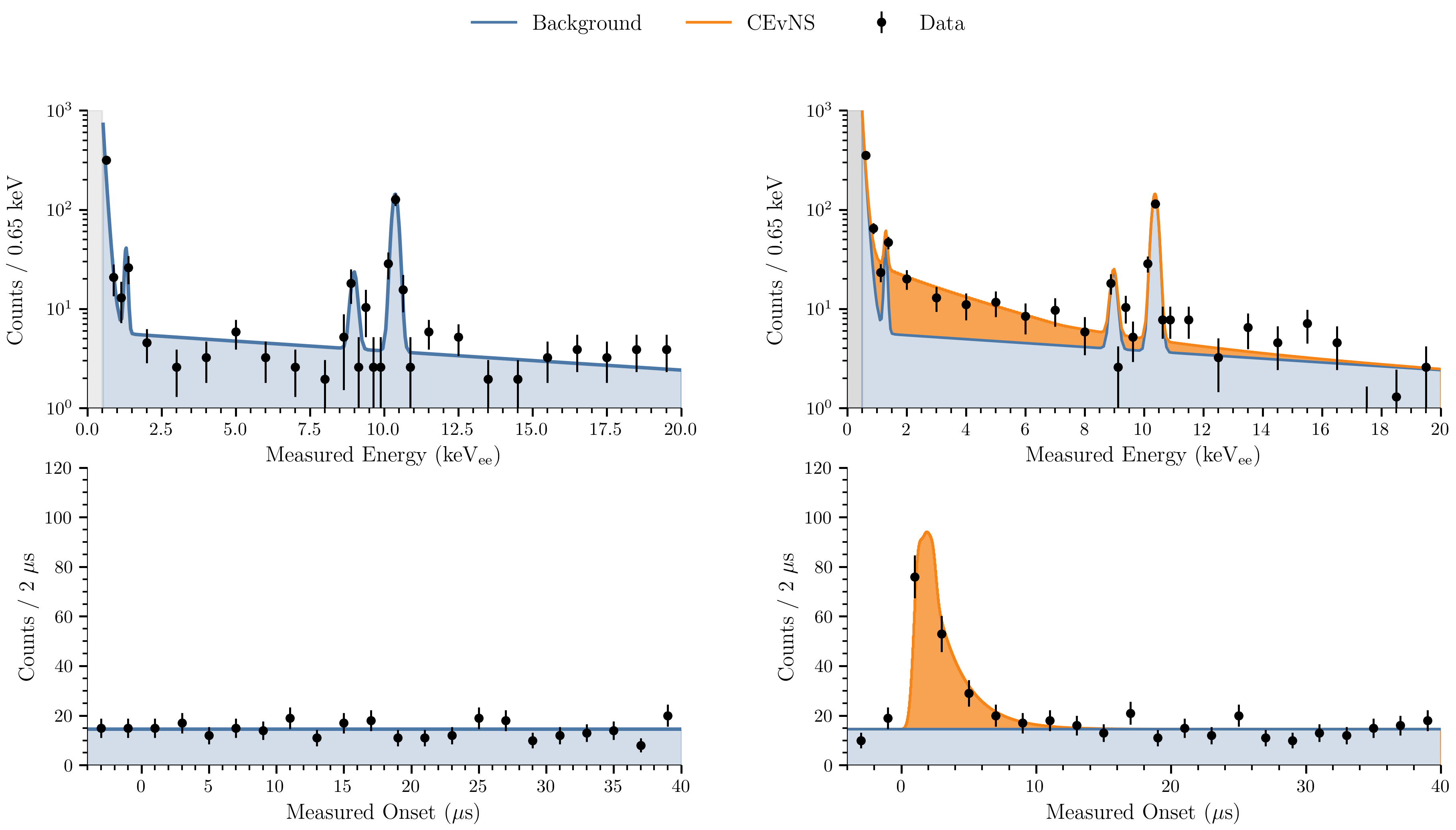}
  \caption{Best-fit maximum likelihood estimates of the signal (orange) and background (blue) distributions projected in energy and time for both off-beam and on-beam data. The energy region of interest is from [0.5, 20]\,keV\(_{\text{ee}}\), while the time region of interest is from [-4, 40]\,\(\mu\)s with respect to an external beam trigger such that the neutrino arrival is expected to begin at -160\,ns. Calibration data, simultaneously fit to constrain the parametrization of the background, is not shown here.}
  \label{fig:results}
\end{figure*}

Among events that pass all data-quality selection, there are two main categories of background events to consider: steady-state backgrounds and beam-related backgrounds. 
Neutron flux measurements in NA, together with dedicated MCNP simulations~\cite{COHERENT:GePRL}, predict a total of (\(2.2\pm0.6\)) beam-related background events in the region of interest. This amounts to \(<2\%\) of the expected signal, and therefore is negligible at the precision of this measurement. This background is not explicitly modeled. The dominant background contribution arises from steady-state sources. The off-beam dataset provides an approximately equal exposure to the on-beam dataset and is a measure of the steady-state background under identical conditions. 

After applying data-quality selections, between 42 and 152 total background events remain per detector within the full energy and timing ROI of the off-beam dataset. The timing distribution of events in all detectors is consistent with a uniform rate, as expected given the low overall event count. The combined energy and time projections of the off-beam data are shown in Figure~\ref{fig:results} on the left.

The steady-state background for each detector is described using an analytic model to capture both the rate and spectral shape. The timing distribution is modeled as uniform, while the energy spectrum is described by two exponential continua representing the low- and high-energy components, together with three Gaussian peaks corresponding to the \(^{\text{71}}\text{Ge}\) K and L shell electron-capture peaks and the \(^{\text{65}}\text{Zn}\) K shell electron-capture peak, which were found to be the most prominent low energy peaks.
Given the low number of steady-state background events, a simple analytic model adequately parametrizes the observed distributions, and any potential deviations from this model are small compared to the statistical uncertainty on the amplitude of the background. The background probability density function is factorized in energy and time. 

Higher-statistics, self-triggered steady-state background data collected during SNS outages are used to constrain spectral features of the analytic background model above an energy threshold of 2\,keV\(_{\mathrm{ee}}\), where the self-triggered data has a trigger efficiency of 100\,\%. Below this threshold, in the energy range [0.5, 2.0]\,keV\(_{\mathrm{ee}}\), the amplitude and shape of the low-energy steady-state background are constrained solely by the off-beam data, which share the same external triggering scheme as the on-beam dataset.

\paragraph*{Signal Prediction.---}

The total accumulated exposure after all operational and analysis dead-times was measured to be 23.65 GWhkg, corresponding to \(4.68\times10^{22}\) protons on target.
The expected CEvNS recoil energy spectrum is calculated assuming the standard-model cross section. The Klein--Nystrand~\cite{Klein:1999qj} nuclear form factor model is adopted with a neutron diffraction radius of \(R_n = 5.006\pm0.052\)\,fm, which has the effect of suppressing the higher energy recoils. 
Expected nuclear recoils in germanium from CEvNS are translated into an ionization yield using the standard Lindhard model~\cite{linhard}, with the predicted free parameter \(k=0.157\) and an assumed uncertainty of \(0.004\) derived from independent measurement in~\cite{Bonhomme:2022lcz}. The analysis threshold of 0.5\,keV\(_{\text{ee}}\) ionization energy corresponds to a minimum nuclear recoil observable of \(\sim2.5\)\,keV\(_{\text{nr}}\). At this threshold, differences between ionization yield models are small, contributing a $0.7\%$ systematic uncertainty
to the cross section measurement. The distribution of CEvNS events in time follows the the distribution of neutrino production at the SNS.

The CEvNS recoil spectrum is smeared for each detector in both energy and time following simulated drift-time distributions matched to characterization data~\cite{COHERENT:GePRL} and measured energy and onset-finding resolution to produce two-dimensional probability density functions in observable energy and time for each detector. The resulting total standard-model expected CEvNS count is \(124\pm13\), with similar exposure in each detector. The systematic uncertainties in the signal expectation are summarized in Table~\ref{tab:systematics}. The systematic uncertainty is dominated by the neutrino flux normalization, and uncertainties in the spectral shape are sub-dominant at the single percent level.

\begin{table}[t]
\caption{Summary of systematic uncertainties on the expected CEvNS signal rate.
Contributions are quoted as relative uncertainties on the total normalization.}
\label{tab:systematics}
\begin{ruledtabular}
\begin{tabular}{lc}
Source of uncertainty & Contribution (\%) \\
\hline
Neutrino flux normalization     & 10.0 \\
Baseline distance               & 0.5 \\
Energy calibration              & 0.8 \\
Active detector mass             & 2.0 \\
Nuclear form factor              & 0.8 \\
Germanium quenching factor       & 0.7 \\
\hline
Total (quadrature sum)           & 10.3 \\
\end{tabular}
\end{ruledtabular}
\end{table}

\paragraph*{Results.---}

An unbinned, two-dimensional maximum-likelihood analysis is performed independently for each detector to extract CEvNS signal counts. For each detector, the likelihood is constructed as a simultaneous fit to the on-beam dataset and auxiliary background-constraining datasets, including off-beam data and high-statistics calibration data. The signal normalization is treated as a free parameter, while the steady-state background is described by a parametric model whose shape and rate are constrained by the auxiliary datasets and profiled in the fit.

The fit is performed over the full analysis window in energy and time for each detector separately.
Figure~\ref{fig:results} shows the off-beam data, on-beam data, and the best-fit results for all detectors summed.

The per-detector likelihoods are then combined in a global fit to the CEvNS cross section, yielding a total of \(124^{+14}_{-12}\) signal events, consistent with the standard-model expectation within \(1~\sigma\). A simultaneous fit across detectors was performed to determine the total flux-averaged cross section, with the neutrino flux normalization treated as a nuisance parameter with a Gaussian prior of 10 percent. The best-fit flux-averaged cross section normalized to the SM expectation of \(5.9\times10^{-39}\,\mathrm{cm}^2\) yielded a value \(1.00 \pm 0.14\).

\begin{figure}
    \centering
    \includegraphics[width=1.0\linewidth]{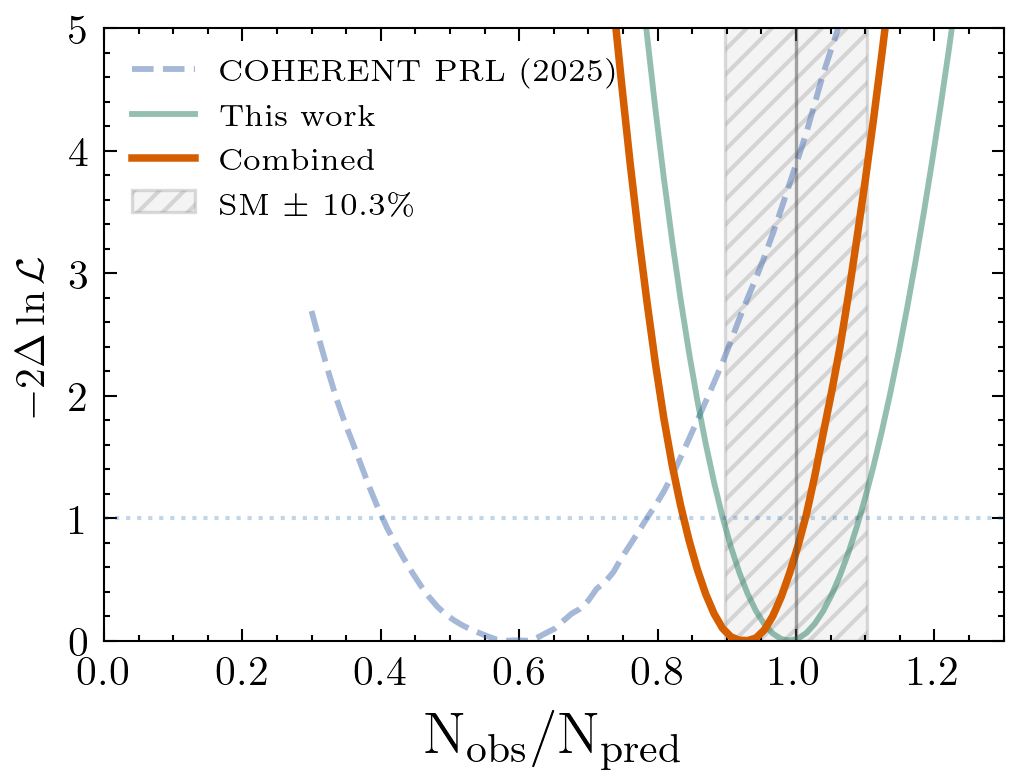}
    \caption{Profile likelihood as a function of the ratio of observed to predicted event rates. The blue curve shows the likelihood curve from Ref.~\cite{COHERENT:GePRL} and green curve shows this work, while the orange curve corresponds to the combined result. The shaded band indicates the standard-model expectation and its associated 10.3\% uncertainty.}
    \label{fig:combined}
\end{figure}

The results reported here are combined with the first Ge-Mini CEvNS result in Ref.~\cite{COHERENT:GePRL} by summing the corresponding profile likelihoods.
Figure~\ref{fig:combined} shows the resulting profile likelihood for the ratio of the observed signal yield to the standard model. The combined result is consistent with the standard-model prediction within \(1\sigma\) and the combined statisical uncertainty is \(9\%\). While the systematic uncertainties for both data sets are comparable, the increased exposure and lower analysis threshold led to a significant reduction of the statistical uncertainty from 30\% to less than 10\%.

The distinct timing structure of the prompt muon-neutrino component and the delayed electron- and antimuon-neutrino fluxes enables sensitivity to flavor-dependent CEvNS rates. 
This dataset is interpreted in the context of non-standard neutrino interactions (NSI) affecting vector neutrino–quark couplings~\cite{Grossman:1995wx, Scholberg:2005qs, Liao:2017uzy}. 
In the presence of a heavy mediator, the coherent scattering cross section is modified through effective flavor-dependent coupling strengths, parameterized by coefficients \(\epsilon^{\alpha \beta}_q\), where \(\alpha, \beta\) denote the neutrino flavor and q the quark type.

\begin{figure}
    \centering
    \includegraphics[width=1.0\linewidth]{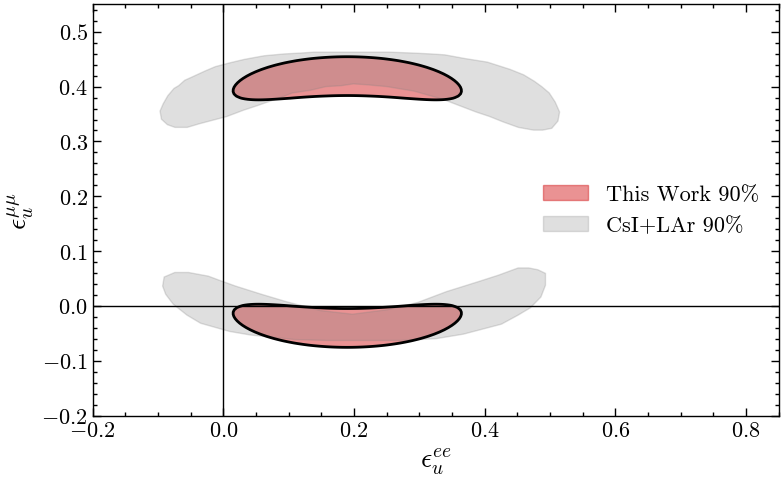}
    \caption{90\% confidence intervals in the \((\epsilon_u^{ee}, \epsilon_u^{\mu \mu})\) plane for previous COHERENT results using the CsI~\cite{COHERENT:CsIPRL} and Ar~\cite{COHERENT:2020ArPRL} detectors in gray~\cite{DeRomeri:2022twg}, and this work in red. The vertical and horizontal lines are the standard model values.}
    \label{fig:nsi}
\end{figure}

Using the same global likelihood framework, constraints are derived by profiling over the standard-model CEvNS normalization and experimental nuisance parameters. Figure~\ref{fig:nsi} shows the resulting confidence regions in the \((\epsilon_u^{ee}, \epsilon_u^{\mu \mu})\) plane.

\paragraph*{Summary and outlook.---}

The COHERENT Ge-Mini detector array at the Spallation Neutron Source has measured the coherent elastic neutrino-nucleus scattering cross section with improved precision, making it the collaboration's first measurement with a dominant systematic uncertainty. An unbinned, two-dimensional maximum likelihood analysis was used to extract the CEvNS signal while profiling a parametric model of the steady-state background, yielding \(123^{+14}_{-12}\) events. 
The timing and energy distributions of the neutrino flux also enable constraints on vector neutrino–quark couplings in the context of non-standard neutrino interactions, yielding more stringent confidence regions in the (\(\epsilon_u^{ee}, \epsilon_u^{\mu \mu})\) plane. 
These results represent the most precise CEvNS measurement to date, and future improvements in the SNS neutrino flux normalization~\cite{COHERENT:2021xhx} will enable even more precise measurements including future Ge-Mini measurements.

\paragraph*{Acknowledgements.-}
The COHERENT Collaboration acknowledges the generous resources provided by the ORNL Spallation Neutron Source, a DOE Office of Science User Facility. Laboratory Directed Research and Development funds from ORNL also supported this project. We acknowledge support from U.S. Department of Energy Office of Science and the National Science Foundation.  
The Ge-Mini array detectors and hardware were acquired through National Science Foundation Major Research Instrumentation Award Number 1920001.
We also acknowledge support from the Alfred P. Sloan Foundation, the Consortium for Nonproliferation Enabling Capabilities, and the Korea National Research Foundation (No. NRF 2022R1A3B1078756). This research used the Oak Ridge Leadership Computing Facility, which is a DOE Office of Science User Facility. We also acknowledge support from Ministry of Science and Higher Education of the Russian Federation, Project ``Studying physical phenomena in the micro- and macro-world to develop future technologies," FSWU-2026-0010.
We thank Mirion Technologies, Meriden CT, USA for guidance and support with the detectors.

\nocite{*}
\bibliography{c3}

\end{document}